\documentclass[pdflatex,sn-mathphys-num]{sn-jnl}

\usepackage{graphicx}
\usepackage{multirow}
\usepackage{amsmath,amssymb,amsfonts}
\usepackage{booktabs}

\graphicspath{{manuscript/}{./}}

\begin{document}
\title[TheseusSyn]{TheseusSyn: Unified Medical Image Synthesis via Context-Conditioned Masked Autoregression}

\author[1]{\fnm{Shibo} \sur{Li}}\email{shibo.li@qmul.ac.uk}

\author*[1]{\fnm{Venet} \sur{Osmani}}\email{v.osmani@qmul.ac.uk}

\affil*[1]{\orgdiv{Digital Environment Research Institute},
    \orgname{Queen Mary University of London},
    \orgaddress{\city{London}, \postcode{E1 1HH}, \country{UK}}}

\abstract{
High-fidelity medical image synthesis has emerged as a promising direction for data-efficient learning and privacy-preserving model development. However, most existing approaches rely on symbolic or label-based conditioning, which constrains generative diversity and limits case-specific contextual fidelity. Inspired by the Ship of Theseus paradox, we propose TheseusSyn, a novel context-conditioned synthesis paradigm that generates images through iterative content replacement.
Our framework employs a multi-codebook discrete tokenizer to convert medical images into sequences of token indices, extracted with a global representation token that serves as a contextual constraint. A masked autoregressive generative model then iteratively masks and predicts tokens conditioned on both local neighborhoods and the global context. By progressively replacing image content until the entire token sequence is renewed, TheseusSyn produces synthetic samples that preserve anatomical coherence and semantic consistency while introducing controlled diversity. We evaluate the proposed framework on multiple datasets from MedMNIST by constructing synthetic datasets with varying masking strategies and training ResNet-18 classifiers exclusively on either real or synthetic data. Experimental results demonstrate that classifiers trained on TheseusSyn-generated images achieve competitive performance on official test sets, indicating that the synthesized data preserve diagnostically relevant features and semantic consistency. Furthermore, systematic analysis of masking strategies reveals their impact on structural fidelity and downstream classification accuracy. These results suggest that context-conditioned masked autoregression provides an effective and generalizable solution for label-free medical image synthesis, offering a scalable alternative to conventional data augmentation and class-conditioned generative models in medical imaging applications.
}


\maketitle

\section{Introduction}\label{sec:introduction}
The demand for high-fidelity synthetic medical data has surged, driven by the scarcity of large-scale, high-quality datasets required to train robust medical models~\cite{babu2025position}. Beyond data augmentation, synthetic generation is an important method to address data imbalance, ensuring generalisability across demographic groups and clinical representation for rare diseases~\cite{mendes2025synthetic}. It also provides a robust response to the tightening constraints of patient privacy and data governance~\cite{giouroukou2025rethinking}.

The rapid evolution of generative frameworks, including Generative Adversarial Networks (GANs)~\cite{armanious2020medgan}, Diffusion Models (DMs)~\cite{graikos2024learned,yellapragada2025zoomldm}, and Autoregressive (AR)~\cite{li2023mage,yuan2024adapting} models, has led to significant progress in medical image synthesis. Current state-of-the-art methods predominantly rely on conditional generation, typically utilizing class labels or text prompts as guidance. However, these paradigms face inherent limitations. Label-~\cite{yuan2024adapting} or text-conditioned~\cite{xu2024medsyn,kim2024controllable} synthesis relies on discrete, artificial symbols to navigate the complex manifold of medical imaging. These methods often target distribution-level characteristics rather than case-specific nuances. Furthermore, class-conditioned methods frequently oversimplify clinical reality by using single-label conditioning, which ignores comorbidities and intersecting pathological factors present in real-world cases. Such models are also rigid, extending them to new labels or additional modalities usually necessitates retraining the entire architecture. For text-to-image synthesis, the challenge lies in the complex alignment between clinical narratives and radiological features, where the final quality is highly sensitive to the structure of the text prompt.

To overcome the constraints of symbolic conditioning, we propose a new paradigm for high-fidelity medical image synthesis that operates independently of human-defined labels. Our goal is to develop a generalizable synthesis framework capable of capturing the complex context of a medical image at a case-specific level. The inspiration for this approach is drawn from the ``Ship of Theseus'' paradox, a classic thought experiment regarding the nature of identity. The myth describes a ship restored part by part until no original components remain. \textbf{If every part of the ship is replaced, is it still the same ship? This creates a state of duality: although its physical constituents change, its essential structure and identity are preserved.} 
We adapt this logic to medical imaging: by treating an existing image as the ``ship'' and selectively replacing its components through masked autoregression, we can generate a novel sample, while preserving its clinical essence and semantics.

In this paper, we introduce TheseusSyn: a unified medical image synthesis framework based on context-conditioned masked autoregression. TheseusSyn consists of three core components: a unified multi-codebook discrete image tokenizer, a context-conditioned masked autoregressive generative model, and a high-fidelity decoder. Given a reference image, the framework discretizes it into token indices---analogous to the wooden planks of Theseus' ship---and extracts a global context representation. By masking and then predicting these tokens conditioned on the global context, TheseusSyn reconstructs the ``ship'' with new components. This approach offers several distinct advantages: (1) \textbf{Modality unification:} by utilizing a context-based representation rather than explicit labels, a single model can generalize across diverse modalities and anatomical regions; (2) \textbf{Case-Specific context conditioning:} being label- and prompt-free, the model uses the image itself as the primary descriptor. This allows for the generation of high-fidelity synthetic images, even in presence of scarce data, such as for rare diseases; (3) \textbf{Fine-Grained controllability:} the masking mechanism allows precise control of the synthetic regions, providing a tunable balance between image fidelity and diversity. We evaluate TheseusSyn framework across 11 distinct image modalities representing 8 different organs, using MedMNIST 2D. Our results demonstrate that the proposed unified paradigm is highly effective to generate high-fidelity synthetic images, consistently improving downstream task performance across varied medical imaging domains.

\begin{figure}[t] 
    \centering
    \includegraphics[width=0.9\textwidth]{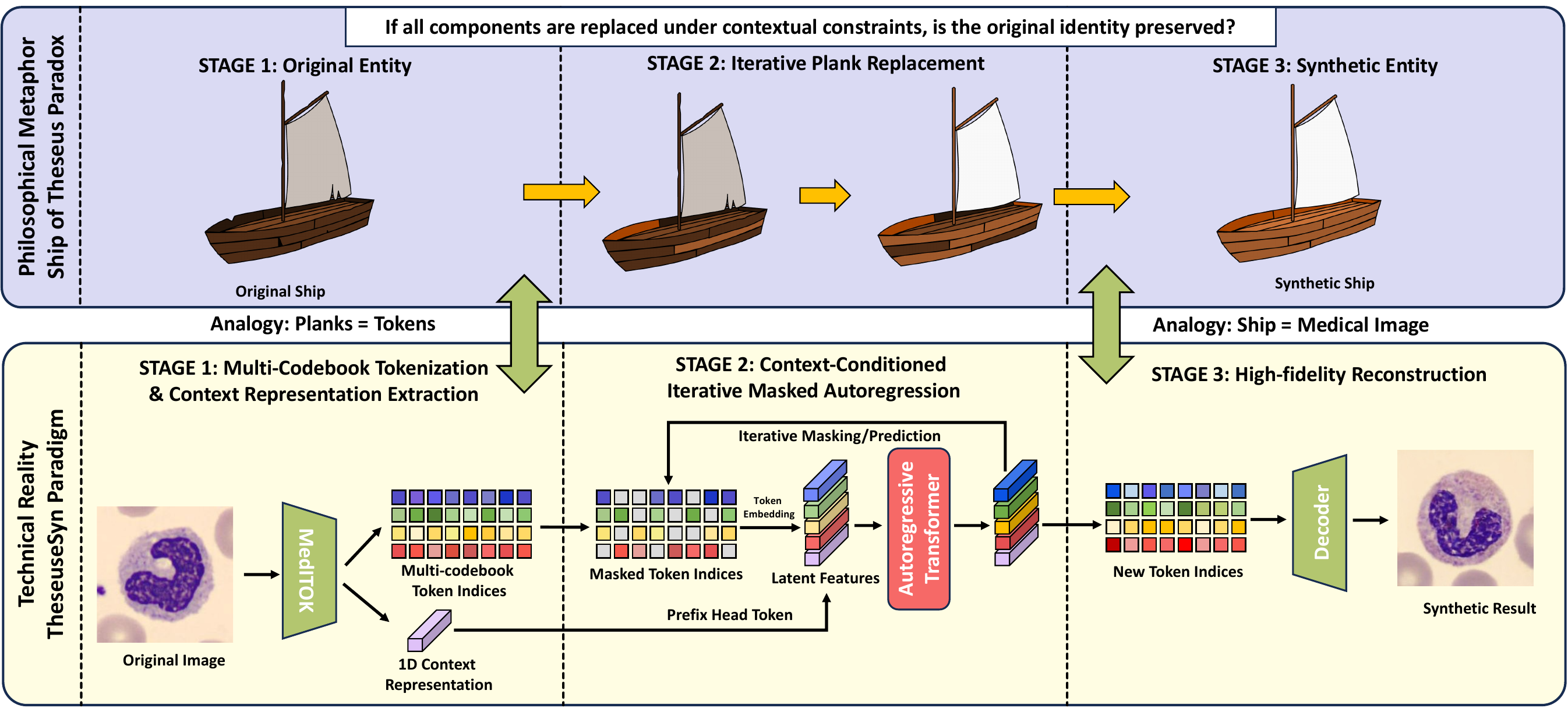}
    \caption{The TheseusSyn paradigm illustrated through the ship of Theseus paradox.}
    \label{fig:paradox_idea}
\end{figure}

\section{Methodology}\label{sec:methodology}
The proposed TheseusSyn framework synthesizes high-fidelity medical images by selectively replacing local anatomical ``components'' while strictly preserving the global clinical semantics of a reference image. As illustrated in the framework overview, TheseusSyn comprises three primary modules: (1) a unified multi-codebook tokenizer to extract discrete image tokens and 1D contextual representation, (2) a context-conditioned masked autoregressive generative model, and (3) a high-fidelity image decoder.

\subsection{Preliminaries: Unified Multi-Codebook Tokenization and Context Extraction}
To discretize the complex manifold of medical images, we employ a unified pretrained medical image tokenizer, MedITok~\cite{ma2025meditok}. Given a reference image $X \in \mathbb{R}^{H \times W \times C}$, MedITok maps it into a sequence of discrete token indices $Z \in \mathbb{Z}^{L \times K}$, where $L$ is the spatial sequence length and $K$ represents the number of parallel codebooks, each with a vocabulary size of $V$. Unlike standard single-codebook quantization, such as VQ-GAN~\cite{esser2021taming}, which forces complete spatial masking, the multi-codebook approach inherently reduces compression loss and introduces a feature dimension to the token indices. This allows for fine-grained partial masking at the sub-token level. Additionally, MedITok~\cite{ma2025meditok} extracts a 1D global context representation $c \in \mathbb{R}^{D}$, where $D$ is the dimension for latent feature embedding, acquired via global average pooling and aligned with vision-language foundation models, to aggregate both the overall clinical semantics and visual contexts.

\subsection{Context-Conditioned Masked Autoregression}
The core of TheseusSyn is a decoder-only Transformer designed to capture both the spatial correlations and the intricate cross-codebook dependencies inherent in the medical images. The objective is to predict masked token indices conditioned on the global context $c$, effectively selecting suitable discrete tokens from the multi-codebook dictionary to rebuild the image.

\subsubsection{Multi-Codebook Masked Token Embedding}
Given the discretized token sequence $Z$, selected positions are replaced with a dedicated $[\text{MASK}]$ index to obtain the corrupted token sequence $Z'$. The indices at each spatial position are embedded independently. For each codebook $k \in \{1, \dots, K\}$, we apply a dedicated BERT-style embedding layer~\cite{devlin2019bert,bao2021beit}. The $K$ resulting embeddings are then concatenated along the channel dimension and passed through a linear projection layer with GELU activation to form a unified $d$-dimensional latent representation for each spatial position.

\subsubsection{Contextual Representation Guidance}
To ensure the autoregressive generation adheres strictly to the reference case's clinical essence, we utilize the 1D context vector $c$ extracted by MedITok~\cite{ma2025meditok}. This vector is projected via a Multi-Layer Perceptron (MLP) and prepended to the latent sequence as a prefix token. This explicitly conditions the self-attention mechanisms within the Transformer blocks, guiding the synthesis of local patches using global clinical semantics.

\subsubsection{Multi-Codebook Prediction Head}
Following the Transformer blocks, the output sequence features are split back into $K$ parallel streams. Each stream employs a dedicated Masked Language Modeling (MLM) layer~\cite{devlin2019bert,bao2021beit} to compute the similarity between the predicted features and the respective BERT word embeddings per codebook, producing $K$ sets of logits. These logits represent the confidence or probability distribution of the discrete tokens for each codebook at each spatial position. During generation, we apply temperature scaling to these logits for parallel token sampling, which allows us to balance synthesis fidelity and diversity. Finally, the sampled sequence of multi-codebook token indices is mapped back into the continuous pixel space using the high-fidelity MedITok decoder to construct the final synthetic image.

\subsection{Training Strategies}
We implement three core training strategies to train a robust, unified model capable of synthesizing diverse medical imaging modalities and tolerating wide-ranging masking ratios (0--100\%), namely: the dual dynamic masking engine ensures resilience to diverse corruption patterns, label-smoothed optimization mitigates overfitting, and a unified sampling paradigm addresses dataset imbalance. We provide a brief description for each of these strategies.

\subsubsection{Dual Dynamic Masking Engine}
To ensure robust cross-dimensional feature learning, we introduce a dual dynamic masking mechanism during training: (1) \textbf{Random Masking:} To force the model to learn both cross-codebook and spatial correlations, individual codebook tokens are masked independently based on a variable mask rate, drawn from a truncated normal distribution; (2) \textbf{Position-level Pruning:} To make the model tolerant to entirely missing regions, we selectively drop all $K$ codebook tokens at randomly selected spatial positions. The pruning ratio is also dynamically sampled from a truncated distribution. This mixture yields variable-difficulty reconstruction tasks, acting as a strong regularizer that stabilizes training and exposes the model to diverse corruption patterns. The resulting binary mask is applied to the discretized sequence $Z$, replacing masked positions with a dedicated $[\text{MASK}]$ token to obtain the corrupted sequence $Z'$. 

\subsubsection{Label-Smoothed Optimization}
In synthesizing medical images with autoregressive models, the ground-truth sequence of token indices is often just one of many clinically plausible combinations. Relying on a standard hard-target cross-entropy loss forces the model to ignore this inherent uncertainty, over-penalizing viable alternative tokens. To address this, we optimize the network using a Label-Smoothed Cross-Entropy Loss~\cite{guo2024cross} with a smoothing factor of $\alpha=0.1$. By replacing hard one-hot targets with a softened probability distribution, this loss acts as a critical implicit regularizer that prevents overconfident predictions, improves probability calibration, and enhances robustness against token noise.

\subsubsection{Unified Training Paradigm}
To support diverse imaging modalities (e.g., X-ray, MRI, microscopy) within a single set of parameters, without allowing large datasets to dominate the optimization, we employ a balanced multi-modality sampler. For a unified model spanning $M$ distinct modalities, each training mini-batch is equally partitioned into $M$ sub-batches, with each sub-batch sampled exclusively from its corresponding modality-specific dataset. This design mitigates distribution skew, ensuring that all modalities contribute equally to the gradient updates at every iteration. Consequently, this improves cross-modality robustness and generalization across heterogeneous clinical domains.
\subsection{Inference Strategies}
During inference TheseusSyn utilizes a MaskGIT-style~\cite{chang2022maskgit} iterative parallel decoding scheme. To enable case-specific controllable synthesis, we define three distinct masking modes: (1) \textbf{Full Masking:} All tokens are masked, and the image is synthesized entirely from scratch guided solely by the extracted global context $c$; (2) \textbf{Partial Masking:} A user-specified ratio of tokens is masked, seamlessly blending generated features with real retained tokens. This can be applied via random masking (independent masking across codebooks) or uniform masking (masking the same proportion of codebooks at targeted spatial locations); (3) \textbf{Theseus-style Iterative Masking:} The model progressively replaces all tokens in the reference image over multiple iterations in analog to the ship of Theseus paradox. For example, if we mask and generate 25\% tokens of the image each iteration, then we need at least 4 iterations to replace all of the tokens while rigorously preserving clinical semantics.

\section{Experiments}\label{sec:experiments}
\subsection{Datasets}
We evaluate \textbf{TheseusSyn} on the MedMNIST v2 benchmark~\cite{medmnistv1,medmnistv2}, a standardized collection of 2D medical images. To assess unification capability, a single model is trained jointly on all 11 MedMNIST datasets spanning eight anatomical regions and imaging modalities, including microscopy (BloodMNIST, TissueMNIST), dermatoscopy (DermaMNIST), fundus photography (RetinaMNIST), retinal OCT (OCTMNIST), pathology (PathMNIST), chest X-ray (ChestMNIST, PneumoniaMNIST), and abdominal CT (OrganAMNIST, OrganCMNIST, OrganSMNIST). For unified learning, the 11 sub-datasets are aggregated into eight categories based on organ type and modality. All images are resized to $224 \times 224$, and the official MedMNIST training, validation, and test splits are strictly followed. Experiments consist of (i) training and evaluation of TheseusSyn, and (ii) downstream task evaluation for synthetic data. Since the two parts cannot use the same training split, TheseusSyn is trained on the MedMNIST training set while the classifiers are trained on the validation set augmented with synthetic data. The test set is used for evaluating both synthesis quality and classification performance.

\subsection{Implementation Details}
The unified multi-codebook tokenizer and high-fidelity decoder are adopted from the pretrained MedITOK model. MedITOK employs $C=8$ codebooks, each with $K=4096$ embeddings, and produces a global 1D context representation of dimension 768. The context-conditioned masked autoregressive generator in TheseusSyn is implemented as a decoder-only Transformer with 16 layers and 12 attention heads, using a latent dimension $d=768$. Token indices from each codebook are embedded using BERT-style embeddings of dimension 96, concatenated across codebooks, and projected to 768 dimensions via a linear layer. The global context vector is projected using a two-layer MLP with an expansion ratio of 4.0. The model is trained using the AdamW optimizer with a learning rate of $5\times10^{-4}$ and label smoothing factor $\alpha=0.1$. Training is performed for 400 epochs with a batch size of 256. To address severe dataset imbalance across the eight categories, as shown in Table~\ref{tab:quantitative}, we apply the unified training paradigm, where each sub-batch is evenly partitioned into eight sub-batches with 64 samples per category. One epoch consists of 500 iterations. The dual dynamic masking engine is set as follows: for per-codebook random masking, the mask ratio is sampled from a truncated normal distribution in $[0.45, 0.95]$ with mean 0.55 and standard deviation 0.25; for position-level pruning, the mask ratio is sampled from $[0, 0.4]$ with the same mean and standard deviation. During inference, a MaskGIT-style parallel decoding strategy is employed with 10 steps and a sampling temperature of 1.0. All experiments are implemented in PyTorch and conducted on a single NVIDIA H200 GPU with 141\,GB memory.

\begin{figure}[t] 
    \centering
    \includegraphics[width=\textwidth,clip]{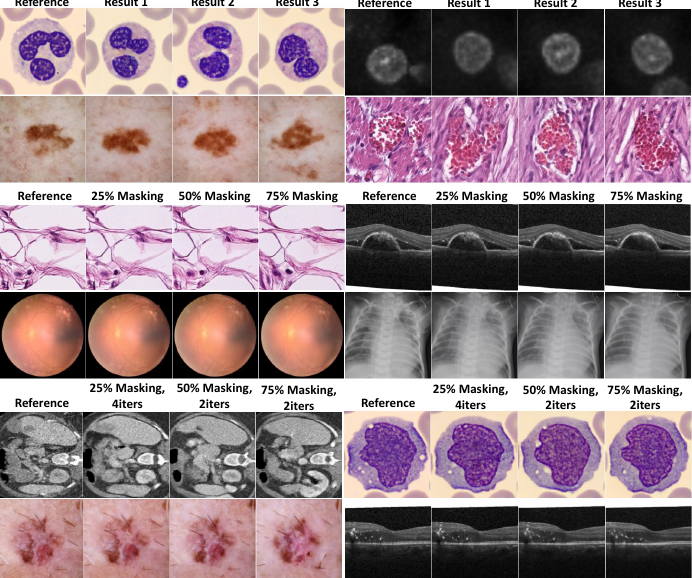}
    \caption{TheseusSyn visual results with different masking strategies on MedMNIST.}
    \label{fig:visual}
\end{figure}

\subsection{Synthetic Image Quality Results}
We quantitatively evaluate synthetic image quality using Fr\'echet Inception Distance (FID)~\cite{heusel2017gans} between real test images and generated samples under different masking strategies (lower is better). As shown in Table~\ref{tab:quantitative}, reconstruction without masking (0\%) yields the lowest FID across all datasets, reflecting the upper bound of image fidelity inherited from the MedITOK tokenizer--decoder. As the masking ratio increases, FID degrades progressively, reflecting the increasing reconstruction difficulty and reduced reliance on original token information. Importantly, partial masking (25\%--75\%) maintains relatively low FID across modalities, indicating that TheseusSyn can generate high-quality case-conditioned variants while preserving global semantics. Under full masking (100\%), generation relies solely on the 1D context representation, resulting in higher FID but still producing anatomically plausible samples across all datasets. Iterative masking substantially improves FID compared to single-step full masking at similar corruption levels, validating the effectiveness of progressive token refinement. Qualitative examples in Fig.~\ref{fig:visual} further illustrate that partial masking preserves context with localized variation, full masking increases diversity under semantic constraints, and iterative masking provides a balanced trade-off between fidelity and diversity. Overall, TheseusSyn demonstrates stable and consistent synthesis performance across heterogeneous medical imaging domains.

\begin{table*}[t]
    \centering
    \caption{Synthetic image quality (FID $\downarrow$) under different masking strategies.}
    \label{tab:quantitative}
    \renewcommand{\arraystretch}{1}
    \renewcommand{\tabcolsep}{1.5mm}
    \resizebox{\textwidth}{!}{%
    \begin{tabular}{ccccccccccc}
    \hline
    \multicolumn{2}{c}{\textbf{FID across MedMNIST 2D}} & \textbf{Blood} & \textbf{Derma} & \textbf{Oct} & \textbf{Fundus} & \textbf{Path} & \textbf{Chest} & \textbf{Organ} & \textbf{Tissue} & \textbf{Avg} \\ 
    \hline
    \multirow{2}{*}{\textbf{\begin{tabular}[c]{@{}c@{}}Dataset \\ Size\end{tabular}}} & Train(TheseusSyn) & 11959 & 7007 & 97477 & 1080 & 89996 & 83176 & 61468 & 165466 & - \\
     & Test(TheseusSyn) & 3421 & 2005 & 1000 & 400 & 7180 & 624 & 17778 & 47280 & - \\ \hline
    \textbf{\begin{tabular}[c]{@{}c@{}}MedITOK\\ Recon\end{tabular}} & 0\% masking & 6.78 & 10.12 & 14.24 & 8.25 & 6.39 & 12.92 & 2.67 & 1.84 & 7.90 \\ \hline
    \multirow{3}{*}{\textbf{\begin{tabular}[c]{@{}c@{}}Partial\\ Masking\end{tabular}}} & 25\% masking & 7.12 & 12.05 & 17.08 & 9.82 & 8.24 & 15.96 & 3.47 & 1.67 & 9.43 \\
     & 50\% masking & 8.05 & 15.37 & 21.07 & 11.72 & 11.38 & 20.75 & 4.81 & 1.71 & 11.86 \\
     & 75\% masking & 10.29 & 20.82 & 28.62 & 15.68 & 17.15 & 29.8 & 7.88 & 3.62 & 16.73 \\ \hline
    \textbf{\begin{tabular}[c]{@{}c@{}}Full\\ Masking\end{tabular}} & 100\% masking & 24.69 & 33.63 & 47.46 & 29.12 & 29.4 & 53.94 & 24.38 & 19.86 & 32.81 \\ \hline
    \multirow{3}{*}{\textbf{\begin{tabular}[c]{@{}c@{}}Iterative \\ Masking\end{tabular}}} & 25\% masking*4 iters & 9.78 & 19.32 & 28.84 & 14.01 & 18.65 & 30.57 & 6.15 & 1.66 & 16.12 \\
     & 50\% masking*2 iters & 10.79 & 22.79 & 31.37 & 16.19 & 20.38 & 34.83 & 7.38 & 2.29 & 18.25 \\
     & 75\% masking*2 iters & 11.90 & 24.45 & 34.18 & 17.87 & 21.51 & 38.29 & 8.96 & 4.03 & 20.15 \\ \hline
    \end{tabular}
    }
\end{table*}

\begin{table*}[t]
    \centering
    \caption{Evaluation of downstream classification using synthetic data.}
    \label{tab:downstream}
    \renewcommand{\arraystretch}{1}
    \renewcommand{\tabcolsep}{4mm}
    \resizebox{\textwidth}{!}{%
    \begin{tabular}{ccccccc}
    \hline
    \textbf{Datasets} & \textbf{\begin{tabular}[c]{@{}c@{}}Train/Test \\ (ResNet-18)\end{tabular}} & \textbf{Accuracy} & \textbf{Real} & \textbf{50\% masking} & \textbf{100\% masking} & \textbf{\begin{tabular}[c]{@{}c@{}}50\% masking\\ *2 iters\end{tabular}} \\
    \hline
    \multirow{2}{*}{\textbf{Blood}} & \multirow{2}{*}{1712/3421} & AUC & \textbf{0.998} & 0.997 & 0.9896 & 0.9956 \\
     &  & ACC & \textbf{0.9658} & 0.9567 & 0.8965 & 0.9433 \\
    \hline
    \multirow{2}{*}{\textbf{Derma}} & \multirow{2}{*}{1003/2005} & AUC & 0.8647 & 0.8517 & \textbf{0.8722} & 0.8649 \\
     &  & ACC & \textbf{0.7147} & 0.6983 & 0.6653 & 0.6888 \\
    \hline
    \multirow{2}{*}{\textbf{Oct}} & \multirow{2}{*}{10832/1000} & AUC & \textbf{0.9884} & 0.9882 & 0.9407 & 0.9793 \\
     &  & ACC & \textbf{0.903} & 0.816 & 0.735 & 0.78 \\
    \hline
    \multirow{2}{*}{\textbf{Fundus}} & \multirow{2}{*}{120/400} & AUC & 0.7123 & \textbf{0.717} & 0.7163 & 0.7091 \\
     &  & ACC & 0.53 & 0.52225 & 0.49 & \textbf{0.5375} \\
    \hline
    \multirow{2}{*}{\textbf{Pathology}} & \multirow{2}{*}{10004/7180} & AUC & \textbf{0.9943} & 0.9937 & 0.9833 & 0.9874 \\
     &  & ACC & \textbf{0.9143} & 0.8985 & 0.8465 & 0.8326 \\
    \hline
    \multirow{2}{*}{\textbf{Chest}} & \multirow{2}{*}{524/624} & AUC & 0.9464 & \textbf{0.9605} & 0.9313 & 0.9584 \\
     &  & ACC & 0.8413 & \textbf{0.8782} & 0.7933 & 0.8141 \\
    \hline
    \multirow{2}{*}{\textbf{Organ}} & \multirow{2}{*}{6491/17778} & AUC & 0.963 & \textbf{0.9667} & 0.9467 & 0.9638 \\
     &  & ACC & 0.7423 & 0.7891 & 0.6891 & \textbf{0.7935} \\
    \hline
    \multirow{2}{*}{\textbf{Tissue}} & \multirow{2}{*}{23640/47280} & AUC & 0.8874 & 0.8728 & 0.846 & \textbf{0.8887} \\
     &  & ACC & \textbf{0.6229} & 0.6044 & 0.532 & 0.593 \\
    \hline
    \multirow{2}{*}{\textbf{Avg}} & \multirow{2}{*}{-/-} & AUC & \textbf{0.9193} & 0.9185 & 0.9033 & 0.9184 \\
    &  & ACC & \textbf{0.7793} & 0.7704 & 0.7060 & 0.7479 \\
    \hline
    \end{tabular}
    }
\end{table*}

\subsection{Downstream Task Evaluation of Synthetic Images}
We further evaluate the task-level utility of synthetic images for classification. For each dataset, we construct synthetic sets from the MedMNIST validation split with the same size as the real set under different masking strategies. A ResNet-18 classifier is trained on either real or synthetic data and evaluated on the official test set. Performance is measured using AUC and accuracy (ACC). Table~\ref{tab:downstream} shows that partial masking-based synthesis (50\%) consistently improves or matches baseline performance trained on real data alone, particularly on challenging data-scarce datasets such as Fundus, Chest, and Organ. In contrast, fully masked samples (100\%) generally leads to degraded performance, reflecting reduced fine-grained fidelity despite increased diversity. Iterative masking recovers part of this performance loss and yields stable improvements in several datasets, highlighting the benefit of progressive token replacement. Overall, synthesis with partially masked or iterative synthetic images achieves competitive AUC and accuracy compared to training on real datasets, demonstrating that the synthetic results generated by TheseusSyn can effectively support downstream tasks.

\section{Conclusion}\label{sec:conclusion}
In this work, we introduced \textbf{TheseusSyn}, a unified medical image synthesis framework based on context-conditioned masked autoregression. Unlike class- or prompt-conditioned paradigms, TheseusSyn directly conditions on intrinsic image context, enabling controllable synthesis without reliance on human-defined annotations. Extensive experiments on MedMNIST demonstrate that TheseusSyn achieves robust image quality under varying masking patterns and produces synthetic samples that retain meaningful clinical features for downstream classification tasks. We believe TheseusSyn potentially provides a scalable foundation for unified medical image generation and offers a practical generalizable pathway toward privacy-preserving data synthesis and balanced learning in heterogeneous clinical domains.

\backmatter

\section*{Declarations}
\subsection*{Funding}
No funding was received for this study.

\subsection*{Competing interests}
The authors declare that they have no competing interests.

\subsection*{Ethics approval}
Not applicable. This study used publicly available benchmark datasets and
involved no new collection of data from human participants or animals.

\subsection*{Data availability}
All datasets analyzed in this study are publicly available as part of the
MedMNIST v2 benchmark~\cite{medmnistv2}. The data can be accessed from the
official MedMNIST Zenodo repository at
\url{https://doi.org/10.5281/zenodo.6496656} and through the project website at
\url{https://medmnist.com/}. No new clinical or patient-level data were
collected for this study. The synthetic images used in the reported experiments
were generated from these public benchmark data using the procedures described
in this article.


\bibliography{ref}

\end{document}